# Current Trends in Physics Enrollment

Samina Masood
Department of Physical and Applied Sciences
University of Houston-Clear Lake

## Abstract

We analyze the current trends in higher education and discuss its impact on physics enrollment in US institutions. The pandemic, lockdowns, unemployment, and healthcare problems have led to unique social and economic conditions. These conditions have modified the latest trends in education. COVID-19 has had great impact on the academic culture due to online teaching and learning methods. We identify some of the key factors including economic problems, changes in job market, modifications in family obligations, physical and mental health conditions, and overall insecurity and uncertainty in life. These key factors are causing a shift in educational preferences. A few recommendations are made to get out of the current dilemma. We use all the data collected by the American Physical Society statistics department [1].

## Identification of the Problem

Physics has never been a very popular subject and has been almost accepted as a low enrollment discipline. Physics departments are relatively small departments and physics programs are smaller programs and have to be continuously watched and special recruitment tools are used to avoid further decrease in enrollment. This enrollment may easily go down, especially in teaching campuses such as community colleges or undergraduate institutes. The decrease in physics enrollment [2] has been a topic of discussion for a long time. The American Physical Society (APS) has been studying this issue and has provided enrollment and graduation data [1] with demographic details.

Physics recruitment strategies and program enrollment issues are emerged from socio-economic standards of educational and cultural values and their direct impact on society cannot be ignored. We therefore have to analyze undergraduate and graduate programs separately with their own problems and issues embedded in the nature of the physics subject and its image. It is known that physics is considered to be a very mathematical subject and has a reputation of an involved discipline which requires intelligent and hardworking people. In this time of easy-going approach towards life and schooling with financial responsibilities, physics does not seem to be the best choice if the job-market is not exceptionally promising either.

Physics is accepted as an involved subject and cannot be studied without personal interest, aptitude, commitment, and dedication. It is considered the mother of technology by scientists as you cannot understand engineering or develop technology without proper physics background. Therefore, introductory physics courses are required for all STEM (Science, technology, engineering and mathematics) majors. However the ratio of physics graduates among all STEM majors is very slim. Figure 1 gives a plot of the ratio of physics graduates among STEM graduates, which is always less than 2.5 percent. We have used ten years of data from APS website to plot this graph. This plot shows the fluctuation in ratio of physics degrees and a recent decrease in this ratio is clearly obvious.

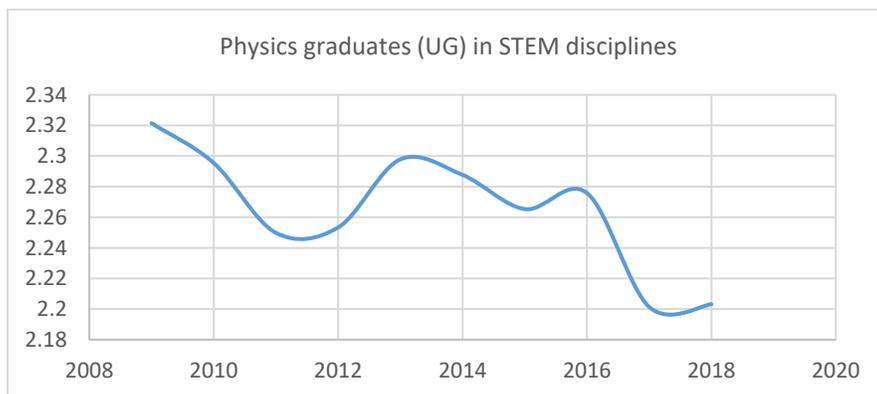

Fig.1: Ratio of Physics graduates as compared to STEM graduates, based on ten year APS data [1]

Traditionally, physics program is fed by its service courses, which included physics for non-science majors, conceptual physics and introductory courses. These courses may have a very large enrollment in big campuses [2]. Introductory courses are offered even if physics degrees are not offered in a given college. Another effective source of funding for physics programs, regardless of low enrollment, has been astronomy and astrophysics courses. These are usually offered by physics programs because standalone astronomy programs are not very common and physics programs offer astronomy and astrophysics courses as well.

On the other hand, undergraduate physics programs have a bigger pool of domestic students and local recruitment strategies contribute significantly. In most community colleges, four year colleges, and even large universities, several introductory physics courses are offered as service courses that may be survey courses for non-science majors. Algebra based introductory physics courses, required by a few disciplines, and calculus based courses are offered for science and engineering majors. Introductory courses, especially the calculus based courses sometimes become recruiting tools for physics majors, if the instructors can succeed to make these courses interesting. Physics courses are needed to run several programs within science and technology disciplines in a university. These programs survive by generating money from service courses because they are needed for all science and engineering disciplines, which are enough to feed these programs. Service courses may have large enrollment as well.

Satellite campuses with MS programs have low graduate enrollment. Ph.D. programs are usually offered in large universities only and do not have such a big issue if they have the teaching assistant and research associate positions. However, theoretical physics degree scholars are mainly international students whereas the applied physics have a regular distribution of ethnic backgrounds in the college. The research projects in applied physics support these programs, especially if they overlap with the engineering disciplines and possible industrial applications.

Funding for graduate study, especially in Ph.D. programs, depends on the federal and state funding policies. However, this support for natural sciences is more challenging. The available funds are essential to support graduate enrollment and financial support for undergraduate studies, which can help in increasing undergrad enrollment as well.

The Physics programs have been slimming down in most cases and it has been discussed several times in literature [3-4]. Improvement in K-12 teaching methods [5-7] and some other methods including modeling [8-11] have been instituted for this purpose. However, all of this work is a part of ongoing efforts towards the attraction of students for physics programs. In this work we have briefly pointed out that now we have to realize the change in the current situation is due to more than two years of the pandemic. We still are not very sure if the pandemic is over or not. The fear of COVID, all type of insecurities, and other uncertainties are still there. However, we are getting used to it. Moreover, the pandemic has changed (at least temporarily) the approach towards life. Current scenarios are different for everybody including the educational institutions, because of the following:

- Heath issues of COVID survivors
- Lockdowns and loneliness
- Layoffs and financial issues
- Change in housing due to unaffordability
- Loss of members of families or friends
- Fear of losing loved ones or emotional shocks and health issues
- Insecurity and uncertainty
- Increasing mental health problems due to stress contributed by all of the above factors

In addition to all these personal issues, virtual teaching and online learning were not easily adoptable educational modes. The virtual communication tools brought in new challenges for older people as compared to younger ones. Virtual communication tools brought changes in the required qualifications in most of the job positions. The lockdown made many people lose their jobs and survival in the job market was a big challenge. Most small industries and jobs related to entertainment, travel, and fashion industry could not survive. Healthcare, food, deliveries, and virtual communication tools became the new trend in the job market. Working from home became a requirement for several jobs instead of a privilege for employees.

The value of human relations and need of families and friends became more valuable while face to face events were replaced by online events and proficiency of technology was required to participate. More people started to live together due to financial needs. On the other hand, more disagreement and intolerance was observed along with the higher divorce rate. All these personal life problems were having a great impact on educational culture, because both students and instructors were getting affected. Due to the lockdowns, parents of young children among educational community had to take care of kids at home, like everyone else. Baby-sitting and helping students in online classes was much more time consuming This adversely affected parents peace of mind and concentration for study.

Stimulus funds and other financial relief packages provided some support. Financial crisis and student loan payment relief actually supported a bit of higher education in USA during the early months of pandemic. But lockdowns, online learning difficulties, irregularities and general impact of the pandemic did not let the educational commitment sustain for a long time. The pandemic has shifted some people towards families and friends and the others lost their interest in education like other worldly matters due to the realization of instability and uncertainty in life. Financial issues took away the affordability of education in some cases and students have to choose families over educational careers. Some students dropped out due to moving back home with families or lack of time. COVID survivors faced health challenges as well as their learning ability and concentration in studies was reduced significantly. All of these issues had an impact on educators and learners as the preference of education moved away from its original position. Many people found almost a new workplace after the educational institutes returned back to face to face delivery mode. Majority of educators and administrators were moved out of the office due to retirement or relocation.

In addition, COVID attracted a lot of interest towards the protection against the pandemic and health and safety took over several other disciplines as funding was moved towards healthcare and safety, SOPs (Safety Operating Procedures), medical equipment, and other basic necessities of life. Mobility decreased and the job market had a growing trend of necessities and development in technology, and facilities as enjoyment efforts were pushed back. Information technology (IT) changed focus to online learning to manage life with virtual connections to avoid the spread of COVID and manage necessities of life by online connections. The entertainment industry, means of transportations, parties, fashion and enjoyment related goods were moved back from the scene temporarily and proficiency with IT was required to survive with normal life instead of special expertise. The focus of study was naturally moved towards the market trends during pandemic. Unfortunately, physics program did not get any popularity during Pandemic. In fact, it lost enrollment because of the above mentioned issues and lack of available time for study.

COVID related data is still being analyzed and we are not fully aware of its impact on mental health. The SARS-COV-II virus and its various mutations are still being studied. However, it is expected to affect mental abilities because its spike protein may anchor wherever receptors are available. However, proper understanding of the virus and its impact on brain cells and related mental health issues are still under-investigation. Long term impact of virus on COVID-19 survivors is not known yet. COVID survivors and families are either worried about all kind of serious after-effects on lungs and other areas of body and live with fears of bad health and others choose to be ignorant. Scientists from all areas of knowledge could contribute [12] to this issue but COVID related topics naturally took over almost everything including genetic study of SARS-COV-II, medical treatment of diseases related to COVID impact, vaccine development, SOPs improvement for long-term use, social and psychological impact of COVID and cultural changes due to the pandemic became popular topics of study. A lot of work is being done to study the impact of COVID-19 on education, pros and cons of online teaching and learning systems [13-14]. However, we do not yet have enough data to conclude the impact of the pandemic. Moreover, these factors have obvious deep-rooted impacts on life, job preferences and students enrollment in various disciplines. But this change was not favorable for physics. Popularity of biology and computer science and engineering was naturally associated with healthcare and pandemic but physics was nowhere near there.

In this situation, we need to change the overall image of physics programs to make it attractive for incoming students and prepare new generations, adequately, for higher education. On the other hand, some immediate steps can be taken to protect current enrollment from further decrease. Some of the recommendations include

- Involvement of faculty in student recruitment, effectively. For this purpose, relevance of physics to daily life and its application to other fields such as computational physics, chemical physics, astrophysics, biophysics, and medical physics should be discussed with incoming students,
- Physics needs to be introduced conceptually and mathematical and computational training should be introduced as tools to learn physics but not to introduce abstract approach to physics.
- Experimental physics in reference to daily life applications promote applied physics more, which makes it attractive in the current lifestyle of the technology era.
- Abrupt changes in fast moving societies come along with various socio-economic pressures for young generations that results in uncertainties, insecurities and confusions. Mental health issues are growing rapidly. Faculty needs to be educated about growing mental health problems and trained to accommodate their emotional needs without compromising on the standard of education.
- The aftereffects of the pandemic on academic culture and direction of its future growth is still not clear. Pandemic is not even over yet. It does not seem plausible to develop future strategies for the post-pandemic era.

In the current situation, we may not do much about future strategies to reduce negative impact of pandemic, which we do not know for sure. However, faculty and administrators are also equally affected by the pandemic and may not be ready to make drastic changes. However, some quick steps can be taken to make a difference. Politics of academia and inappropriate distribution of power may have worse effects under present challenges. Faculty can play pivotal role in protecting low enrollment and assignment of courses and other academic duties should be assigned based on experience and training of faculty and not due to strong lobbying in political institutes. For example:

- Course assignment to faculty should be done based on their training. Some of the faculty members may be trained to teach introductory classes better than advanced courses. They can assign more introductory classes and teach undergraduate core courses.

- Graduate course assignments are needed to be done carefully. Even the core courses should be taught by faculty that have a PhD in the related areas. Theorists can teach most of the core theory courses whereas experimentalists can handle courses in the applied physics better.
- Graduate faculty should be given preference if they are relatively active in research and have the ability to lead research projects. Active researchers can always make graduate courses more interesting and satisfy enthusiastic students.
- Better academic advising is another important factor for the success of a program.
- Proper course selections and the availability of research opportunities is important to attract students who are interested in higher education.
- Faculty assignment should be done based on their training and expertise to get the best out of their skill and expertise so that they can incorporate their research experience in courses and add some updated information from recent research in the subject and make courses more exciting.
- Advising is an art. We need to treat students as if they are very important to us and they feel special and being taken care off. Students' individual needs have to be addressed during advising. Choice of appropriate courses is very important and proper advice is required.

We strongly believe if the above steps are taken and faculty is encouraged to explain in class the importance of physics in terms of its applications to various disciplines and daily life, that enrollment will increase. We have personal experience of students' recruitment from university physics courses. We also have experience of students adding physics as a second major at undergraduate level due to some overlapping research interest such as biophysics and astrophysics ended up getting in to Physics Ph.D. program. Students with undeclared undergraduate majors get in to physics with proper information. A complete change of major from other disciplines is seen as well.